# Laminar boundary layer separation over a fully porous bump

Grace Bridge* and Wen Wu[†]
*Department of Mechanical Engineering, University of Mississippi, University, MS, 38677, USA*

## I. Introduction

The separation of boundary layers over solid surfaces has been extensively studied due to its significant impact on aerodynamic performance and its relevance to the aviation and defense sectors. In contrast, boundary layer separation over porous structures has received comparatively less attention. Examples of such flows include leading-edge separation caused by rime ice — formed by the rapid freezing of small droplets that trap air — on aircraft wings; biomass (e.g., bryozoans and seaweed) accumulation on diverging ship hulls where recirculation may occur; and flow recirculation over oyster reefs, where layers of oysters form a porous region that filters nutrients from the passing flow.

The use of porous materials and active or passive mass transfer through surfaces has emerged as a promising strategy for controlling and mitigating flow separation. The interaction between a fluid flow and a porous surface fundamentally alters boundary layer dynamics compared to traditional smooth or rough surfaces. Porous surfaces enable mass transfer - such as suction or blowing - across the wall, which can significantly modify the velocity profile within the boundary layer. Numerous studies have investigated the impact of porous surface structures on flow separation. For instance, Taamneh et al.[1] found that a porous shell surrounding a solid sphere can delay flow separation, with the effect becoming more pronounced as the permeability increases. However, it also leads to a significant increase in drag. On the contrary, Bhattacharyyaa et al. [2] studied a porous cylinder at very low Reynolds number with modeled pore-flow transportation, and found that drag is slightly reduced by the porous media. Tang et al. [3] simulated flow through and around porous blocks of different height-to-thickness ratios. They observed that the separation region is shortened after the permeability reaches certain thresholds, and so does the drag. In most of these studies, the porous region is modeled rather than resolved in numerical simulations, or bypassed in experimental measurements. While direct measurements of pore-scale flow remain experimentally challenging, high-fidelity numerical approaches capable of resolving flow within porous media are emerging. Recent efforts include Refs. [4–7]. However, studies focusing on non-equilibrium turbulent flows such as flow separation remain rare.

In this work, we use pore-resolved direct numerical simulations to investigate the separation of a laminar boundary layer over a Gaussian bump that is fully porous, formed by packed spheres. The pore flow exhibits rich dynamics, driven by favorable pressure gradients on the windward side of the bump and adverse pressure gradients on the lee side. Preliminary results are presented in this abstract. The full paper will include comprehensive statistics from a more systematic parametric study.

## II. Methodology

### A. Problem formulation

The configuration used for this study consists of a laminar boundary layer developing over a flat plate, on which a porous bump is placed to induce flow separation (figure 1). The two-dimensional profile of the bump surface is described by

$$y(x) = h \exp\left[-\left(\frac{x - x_c}{x_0}\right)^2\right], \tag{1}$$

where $x$ and $y$ are the streamwise and wall-normal directions, respectively. Here, $h = 2.5\delta$ is the bump height ($\delta$ is the thickness of the laminar boundary layer at the inflow), $x_c = 25\delta$ is the location of the bump crest, and $x_0 = 5.735\delta$ sets the extent of the bump on the bottom plate. At the inflow, the Blasius profile of a laminar boundary layer is applied. The Reynolds number, based on the freestream velocity ($U_\infty$) and the inflow boundary layer thickness, is 1000. The spanwise direction is periodic, and a convective outflow is employed. The computational domain is $88.66\delta \times 15\delta \times 12\delta$, in the streamwise, wall-normal, and spanwise ($z$) directions to include the reattaching flow and its recovery, reduce wall-normal blockage, and enable the natural development of three dimensionality.

---

*Master's Student
[†]Assistant Professor. AIAA Member. Corresponding author. Email address: wu@olemiss.edu

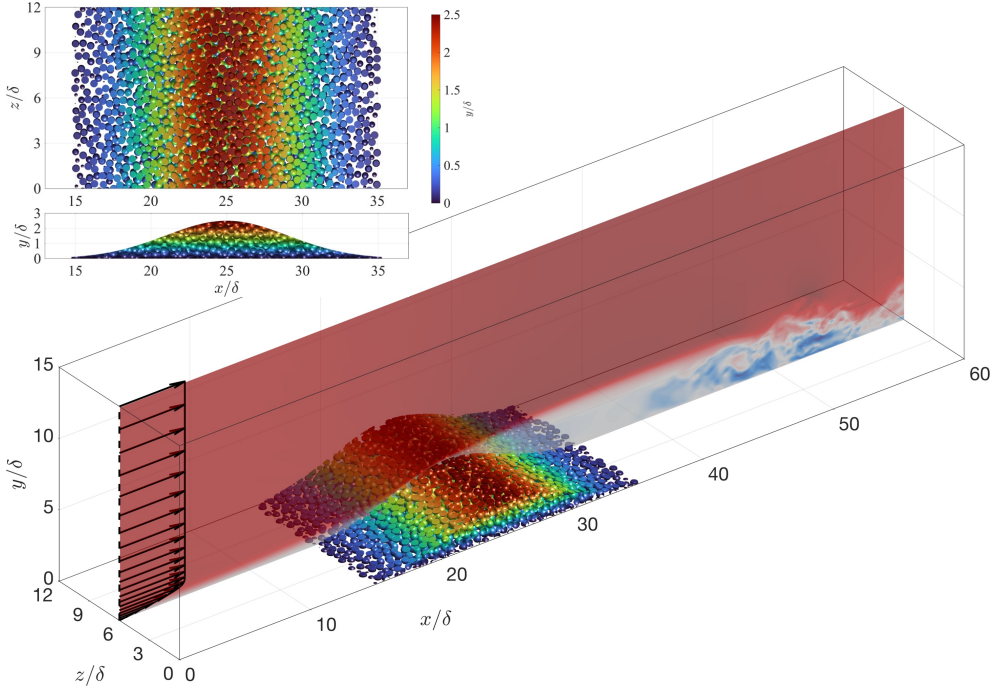

Fig. 1  Schematic of a laminar boundary layer over a porous bump. Only part of the streamwise domain is shown for clarity. The upper left insets show the top and side views of the porous bump, colored by distance from the bottom plate. A contour of instantaneous streamwise velocity is superposed at the mid span for reference.

Comparisons are made between a solid bump and one constructed from stacked spheres (see insets of figure 1). The diameter of each sphere is set to $d = 0.5\delta$, corresponding to 20% of the bump height. To assemble the porous bump, the first sphere is placed with its center at $[x_c, 0, 6\delta]$. Then, additional spheres are added one at a time by attaching each to an existing sphere. Both the choice of the existing sphere which a new one attaches to, and the attachment point are determined randomly. Overlapping spheres, placement with centers below the bottom plate, and filling beyond $|x - x_c| > 2x_0$ are avoided. Periodicity of the spheres is enforced in the spanwise direction. This process continues until no further spheres can be added within the bump region. Once the stacked spheres extend beyond the bump height, they are trimmed along the smooth bump's outline, so that the surface of the resulting porous bump is defined by the cross-sections of the intersecting spheres and aligns precisely with the smooth bump envelope. This approach minimizes surface roughness effects. As a result, the bump forms a porous structure with a smooth, yet transpiring, or partially slippery, surface. The porosity is approximately 0.57, with a small variation in the wall-normal direction (see figure 2). This indicates a relatively loose configuration compared to typical random sphere packing, which generally exhibits porosity in the range of 0.36 to 0.45. This may be due to the random attaching approach lacking particle rearrangement mechanisms.

### B. Numerical method

Direct numerical simulations are preformed by solving the non-dimensionalized incompressible Naiver-Stokes Equations:

$$\frac{\partial u_k}{\partial x_k} = 0; \quad \frac{\partial u_i}{\partial t} + \frac{\partial u_i u_k}{\partial x_k} = -\frac{\partial p}{\partial x_i} + \frac{1}{Re}\frac{\partial^2 u_i}{\partial x_k^2} + f_i. \tag{2}$$

In the above equations, the indices $(i, k) = 1, 2,$ and 3 correspond to the streamwise ($x$), wall-normal ($y$), and spanwise ($z$) directions. The respective velocity components are $u$, $v$, and $w$. Quantities are normalized by the freestream velocity ($U_\infty$) and the boundary layer thickness ($\delta$) at the inflow. A well-validated finite difference code [8, 9] is employed. A second-order Adams-Bashforth scheme is employed for the convective terms while the diffusion terms are discretized using an implicit Crank-Nicolson scheme. The spatial derivatives are computed using a second-order accurate central difference scheme. The Poisson equation is solved with a pseudo-spectral method. The boundary condition on the bump and sphere surfaces is applied by an immersed boundary method. At each time step, the forcing term $f_i$ in the right-hand-side of the NS equation is assigned based on the volume-of-fluid (VOF) of each grid cell to satisfy the no-slip boundary condition [10]. The VOF field is computed in advance based on the positions of the spheres.

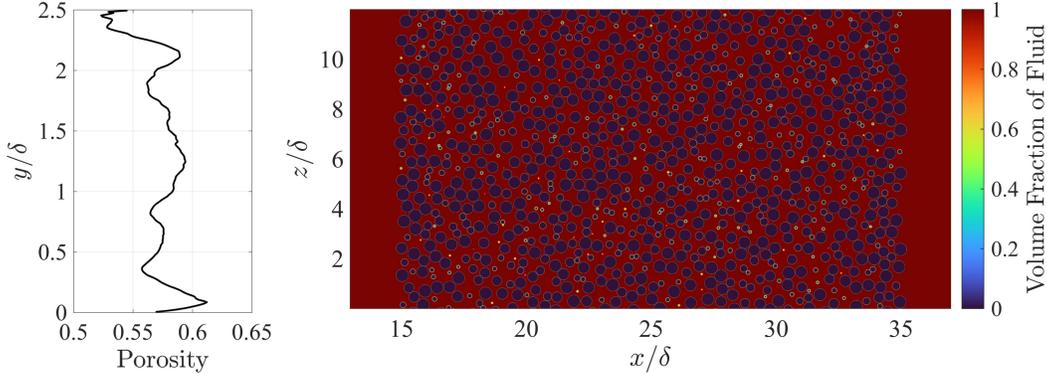

**Fig. 2   Wall-normal profile of the porosity of the porous bump and example of the volume-of-fluid field at the first grid point away from the bottom plate.**

A uniform grid is used in the spanwise direction. For the porous case, the grid is uniform in the wall-normal direction below the bump crest with $\Delta y = 0.01\delta$, and stretched further above. In the streamwise direction, uniform $\Delta x = 0.01\delta$ is maintained for $x/\delta = [15, 35]$, and stretched gradually towards the inflow and outflow. This arrangement ensures that each sphere is resolved uniformly in all three directions by 50 grid points per diameter. The stretching ratio in $x$ and $y$ is less than 2.6%. In total, $3600 \times 407 \times 1200$ (∼ 1.75 billion) grid points are used for the porous case. The maximum $\Delta x$ is $0.105\delta$ near the inflow and outflow, and the maximum $\Delta y$ is $0.17\delta$ near the top boundary. In wall units, the grid resolution in the bump region satisfies $\Delta x^+ < 0.25$, $\Delta y^+(1) < 0.16$, and $\Delta z^+ < 0.35$, while in the far wake of the bump, the maximum values are $\Delta x^+ < 6.2$, $\Delta y^+(1) < 0.3$, and $\Delta z^+ < 0.6$. Since no small solid structure needs to be resolved for the solid bump, this case is computationally cheaper, using $3000 \times 407 \times 600$ (∼ 0.73 billion) grid points arranged similarly to the porous case.

A constant time step $\Delta t = 1.9 \times 10^{-3} \delta/U_\infty$ is used. Statistics are collected over a period of $600\delta/U_\infty$, after the flow reaches a statistically steady state in each case. This sampling duration corresponds to approximately 20 to 30 flow-over times — the convective timescale over which the mean flow travels the extent of the bump. Given the vortex shedding in the bump's wake, this duration spans more than 100 shedding periods, thereby ensuring statistical convergence. Mean quantities are obtained by averaging in time and in the spanwise direction, for which the intrinsic averaging over the fluid region is applied. They are expressed by capital letters or with operator $\overline{(\ )}$ in the following discussion.

## III. Results

### A. Separating shear layer features

The porous bump strongly alters the separating flow, as shown in figure 3. We present the instantaneous wall-normal velocity here to highlight the direction of the pore flow, as well as the large-scale spanwise rotational motion formed by ensembles of smaller eddy structures as the separating shear layer breaks down. In the porous-bump case, the separating shear layer clearly persists much longer. As the favorable pressure gradient strengthens, fluid on the windward side of the bump is driven into the porous medium. This pore flow then traverses the porous structure and emerges as a strong upward and outward flux near the bump's crest. Farther downstream along the bump ($x/\delta \sim 28$–$35$), the pore flow weakens. This attenuation is likely due to the absence of pressure recovery on the leeward side, which appears to choke the flow through the pores.

In the mean sense, the separation region downstream of the bump is increased by 48% in the porous case relative to the solid one (see Figure 4). The reattachment point is delayed from $48.0\delta$ to $56.4\delta$. Here, reattachment is on the flat surface and defined as the location where $C_f = 0$ and $dC_f/dx > 0$. Note that the reattachment in the laminar separation studied here relies on the formation and decay of the roller vortices. Moreover, the onset of separation — defined here as the point where the streamline closest to the bump surface, without penetrating it, detaches — is shifted upstream, from $x/\delta = 25.5$ (just past the crest) to $x/\delta = 23.3$ (before the crest).

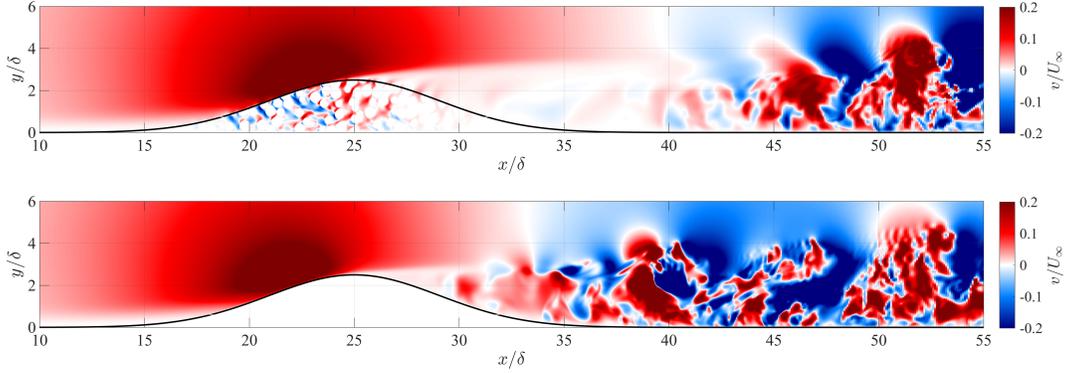

Fig. 3    Contour of instantaneous wall-normal velocity.

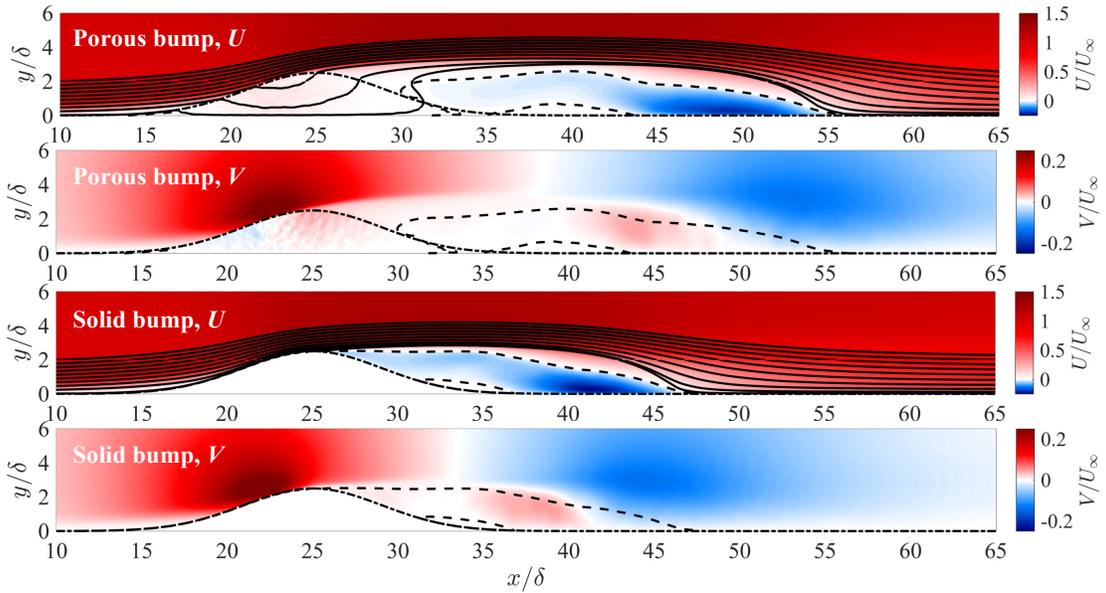

Fig. 4    Comparison of the mean streamwise and wall-normal velocity between the solid and porous bumps. Solid lines in the $U$ contours, selected streamlines starting from $x/\delta = 10$, $y/\delta \in [0.02, 2]$. Dashed line, contour of $U = 0$. Dotted-dash line, outline of the bump surface.

The reduced reverse flow in the front portion of the recirculation region - caused by the added momentum from fluid penetrating through the porous medium - weakens the separating shear layer. This effect is evident in the mean velocity contours, which show a decreased velocity gradient across the shear layer. As a result, the formation of roller vortices is weakened. This is further supported by the turbulent kinetic energy (TKE) contours in Figure 5, where the high-TKE region in the porous case appears farther downstream and persists longer — extending about $20$–$25\delta$, compared to approximately $15\delta$ in the smooth case. These statistics reinforce the observations from the instantaneous flow fields in Figure 3: the roller vortices not only form farther downstream but also remain coherent for a longer duration. This suggests that the flow penetrating through the bump does not introduce strong perturbations to destabilize the separating shear layer but instead plays a more significant role in reducing the mean shear.

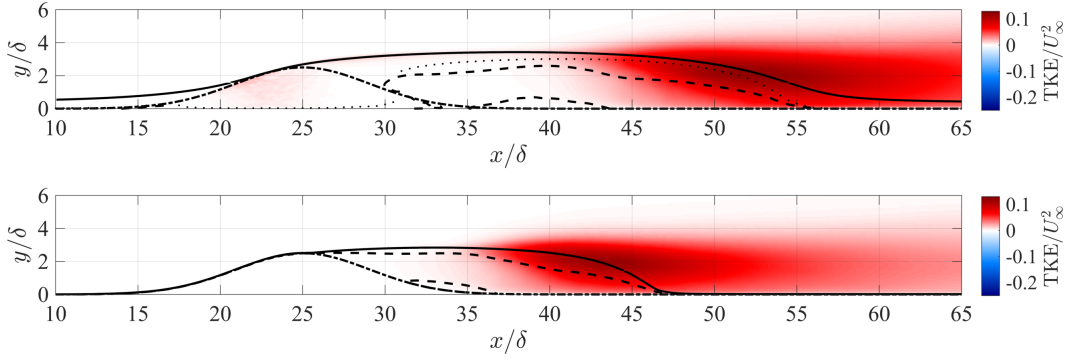

**Fig. 5   Comparison of turbulent kinetic energy (TKE). Solid line: separating streamline. Dotted line: reattaching streamline (coincides with the separating streamline in the smooth case). Dashed line: contour of $U = 0$. Black dash-dotted line: outline of the bump surface.**

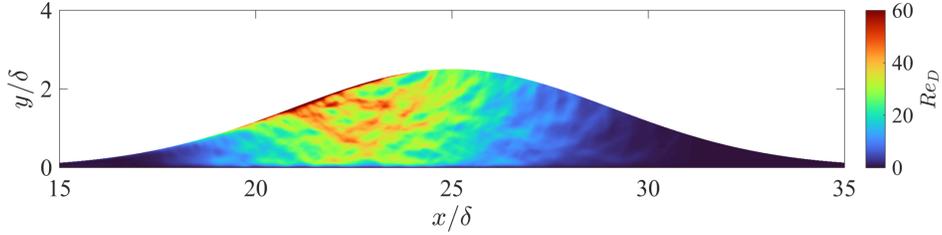

**Fig. 6   Pore flow Reynolds number based on the sphere diameter and pore flow total velocity.**

### B. Pore flow features

Based on the mean streamlines and velocity contours, the pore flow inside the porous bump exhibits the following characteristics: firstly, there is no mean flow reversal within the bump; instead, the flow is primarily driven by a pressure difference caused by the insufficient pressure recovery on the lee side. The peak pore flow velocity magnitude is $0.227 U_\infty$, and the Reynolds number based on the sphere diameter and the pore flow velocity is below 100 (fig. 6). Therefore, the pore flow is viscous-dominated with the close packing of the spheres suppressing the vortex shedding around the spheres.

Secondly, under the favorable pressure gradient on the windward side, the pore flow moves toward the bottom wall, then turns away from it and exits through the lee-side surface of the bump, rejoining the main flow. The location of zero wall-normal velocity ($V = 0$) within the bump is slightly upstream of the bump crest. This is further supported by the measurement of the mean bump-normal velocity taken locally along the surface of the porous bump (figure 7). Clearly, the fluid begins to exit the bump near the wind side vicinity of the bump crest. Then, the escaping 'jet' is most prominent at shortly after the crest. Such flow impinges on and deflects the separating shear layer upward as shown by the streamlines.

The total mass flux through the bump, calculated as the integral of the bump-normal velocity on either the entering or exiting region, is measured to be $0.149\ U_\infty \delta$ per unit span. The basin of this mass is the near-wall flow within $y/\delta < 0.432$ in the incoming boundary layer, accounting for approximately 22.4% of the total mass flux within the boundary layer ($y < \delta$).

## IV. Conclusion

Direct numerical simulations of a laminar boundary layer over a fully porous bump are conducted to investigate pore-scale flow behavior under pressure gradients and flow separation. The pore flow is fully resolved. Compared to the flow over a solid bump, separation is enhanced and reattachment is delayed. On the windward side, the favorable pressure gradient drives near-wall fluid into the porous bump, which is then ejected by the high back-pressure on the lee side. The exiting flux peaks near the crest of the bump, generating a quasi-upward cross-flow jet that gently lifts the separating shear layer. The primary cause of the extended separation region is the reduction in mean shear across the separating shear layer, as the reverse flow is weakened by the added momentum from the exiting pore flow. This, in turn, delays the formation of roller vortices, which play a critical role in reattachment for the laminar flow considered here.

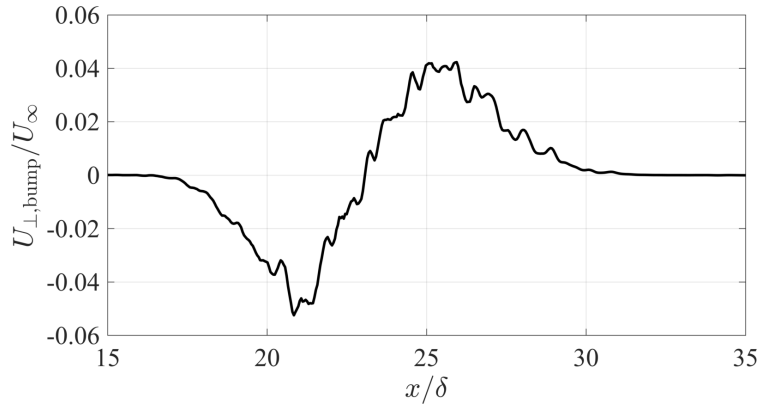

Fig. 7  Profile of the velocity component normal to the porous bump surface. Negative values indicate flow entering the porous bump; positive values indicate flow exiting it. The streamwise center of the bump is at $x/\delta = 25$.

## Acknowledgments

The authors acknowledge the support from the Air Force Office of Scientific Research (Award No. FA9550-25-1-0033), monitored by Dr. Gregg Abate, as well as the National Science Foundation (Award No. 2235036) monitored by Dr. Ronald Joslin. Computational support was provided by the San Diego Supercomputing Center via the ACCESS program funded by the U.S. National Science Foundation.